\documentclass[aps,prl,reprint,superscriptaddress]{revtex4-1}

\usepackage{graphicx}
\usepackage{amsmath}
\usepackage{gensymb}
\usepackage{float}
\usepackage{mathtools}
\usepackage{braket}
\usepackage{romannum}

\usepackage{color}

\begin{document}

\title{Non-local spin valve interferometer: Observation of superconducting spin current and Aharonov-Anandan's non-adiabatic geometric phase}

\author{Ning Lu}
\email{ninglu@physics.sc.edu}
\affiliation{Department of Physics and Astronomy and the SmartState Center for Nanoscale Physics, University of South Carolina, Columbia, South Carolina 29208, USA}
\author{Jian-Sheng Xia}
\affiliation{National High Magnetic Field Laboratory, University of Florida, Gainesville, Florida 32611, USA}

\date{\today}

\begin{abstract}
An electron interferometer was designed and fabricated via a normal metal/insulator/ferromagnet non-local lateral spin valve with a ring-shaped normal metal/insulator spacer, and spin current interference was observed. A very high spin signal of 200 m$\Omega$ was found in a device with 2 $\mu$m injector-detector distance and magnetic field swept parallel to the plane. With a perpendicular magnetic field sweep, a Hanle effect measurement showed both spin precession and $h/e$ oscillation. Because of the non-adiabatic nature of the precessing spins at low fields as they traverse the normal metal ring, this is an experimental observation of Aharonov-Anandan's non-adiabatic geometric phase. In addition, our observation of identical spin resistance for normal and superconducting Aluminum is inconsistent with theoretical predictions based on the quasiparticle picture. To explain the superconducting spin current we suggest that spin triplet Cooper pairs may exist in thin films of Aluminum for direct spin injection.
\end{abstract}

\pacs{}


\maketitle

The Aharonov-Bohm (AB) effect was first proposed in 1959 and experimentally observed in 1985 in a mesoscopic Au ring by electrical quantum phase coherence detection of $h/e$ oscillations during magnetic field sweeps at very low temperature\cite{1, Webb}. The wavefunction phase change due to the AB effect was later reformulated by Berry as a special case of his Berry phase in 1985\cite{Berry_1984, Anandan_1992}. Berry considered an adiabatic geometric phase in which the quantum system is always in an instantaneous eigenstate and whose Hamiltonian changes adiabatically around a closed circuit in parameter space\cite{Berry_1984, Bitter:1987aa}. Aharonov and Anandan (AA) later proposed a more generalized geometric phase for non-adiabatic evolution where the quantum state need not be an eigenstate of the Hamiltonian\cite{Anandan_1992, Aharonov:1987aa}.

In a mesoscopic metal ring structure with interaction between spin and external magnetic field included, the quantum adiabatic approximation is valid for eigenstates in which the spin direction is parallel or antiparallel to the magnetic field and for fields at which the time it takes the electron to transit the ring is much larger than the period of Larmor precession\cite{Stern:1992aa,Frustaglia:2001aa,Hentschel:2004aa}.

In this paper we design and fabricate, for the first time, a non-local lateral spin valve (NLSV) electron interferometer with such a ring structure. We experimentally observe AA's non-adiabatic geometric phase by measuring both the coherent $h/e$ oscillation and spin precession. Spins were injected aligned in plane with the ring and the magnetic field was perpendicular to it. The field was low to ensure that the precession time was longer than the transit time of the ring, resulting in non-adiabatic transport, and therefore making this an AA phase measurement rather than a Berry phase measurement.

\begin{figure}[htp]
\centering
\includegraphics[scale=0.08]{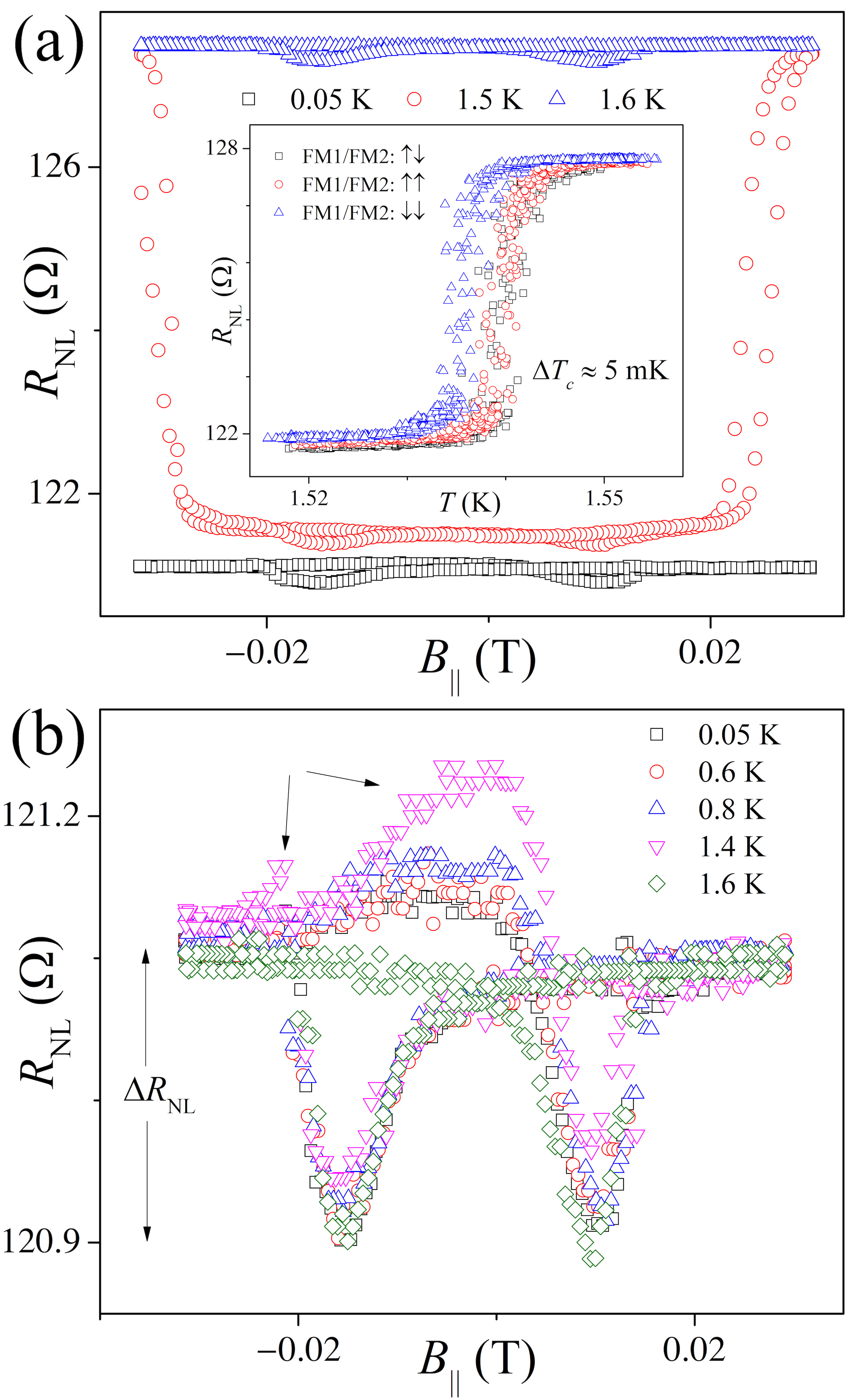}
\caption{\label{fig:noquasi} Co-existence of spin current and superconductivity. (a) Non-local resistance curve with $B_\parallel$ swept between $\pm 0.03$ T at the normal state (1.6 K), superconducting state (0.05 K), and at 1.5 K with a critical field of $B_c = 0.025$ T. Inset: $T_c$ shift for different injector and detector configurations with quasi-static change of temperature at $B_\parallel = 0$. The superconducting transition temperature depends on the degree of spin polarization\cite{Miao:2007aa}. If this is controlled primarily by the injector, we would expect $T_c$ for the the $\uparrow\downarrow$ and $\uparrow\uparrow$ states to be equal but different than for the $\downarrow\downarrow$ state, consistent with our observations. (b) Non-local resistance at various temperatures below and above $T_c$ and at at 1.6 K, after removing the offset due to the superconducting transition. $\Delta R_\mathrm{NL}$ does not change with temperature either below or above $T_c$, which is inconsistent with the theoretical predictions of several orders of magnitude change below $T_c$ based on the quasiparticle picture. In addition, arrows point out a localized increase of $R_\mathrm{NL}$ below $T_c$ when FM1 and FM2 are in the negative parallel configuration. The effect increases as the temperature approaches $T_c$ from below and disappears after the transition. Every $10^{th}$ data point shown for clarity.}
\end{figure}

Additionally, we observe that the spin signal does not depend markedly on temperature in the superconducting state, in contrast to what the quasiparticle picture suggests\cite{Takahashi:2003aa}. We also observe a localized increase in non-local resistance when both the detector and injector are in the negative parallel configuration and the temperature is below $T_c$ (see figure \ref{fig:noquasi}). Finally, spin triplet Cooper pairs are suggested to explain the superconducting spin current.

A NLSV is an effective tool to generate and detect spin polarized electrons in mesoscopic metal and semiconductor systems. In Hanle effect measurements, a magnetic field perpendicular to the plane generates spin precession with a spin direction in plane while electrons diffuse from injector to detector\cite{_uti__2004, Garzon:2005aa, Jedema_2002, Tombros_2007, Fukuma_2011, Han:2010aa, Lou_2007, Dash_2009, Han_2014}. During diffusion, spin can relax via ordinary momentum scattering with spin-orbit coupling\cite{_uti__2004}, i.e. the Elliot-Yafet mechanism. Momentum scattering in diffusive metal films is typically caused by impurities\cite{Bergmann_1984}. Such scattering is elastic and therefore does not break the phase coherence of the electron wave function. Electron-phonon inelastic scattering is the main cause of phase decoherence. But at low temperatures, this is suppressed, thereby enabling the observation of interference effects like those due to weak localization and AB oscillations even in disordered diffusive metals\cite{Webb, Bergmann_1984}. However, in the original AB effect measurement in Au rings, spin was not considered, nor was the spin-magnetic field interaction from the field that penetrates the ring itself\cite{Webb}.

\begin{figure}
\centering
\includegraphics[scale=0.06]{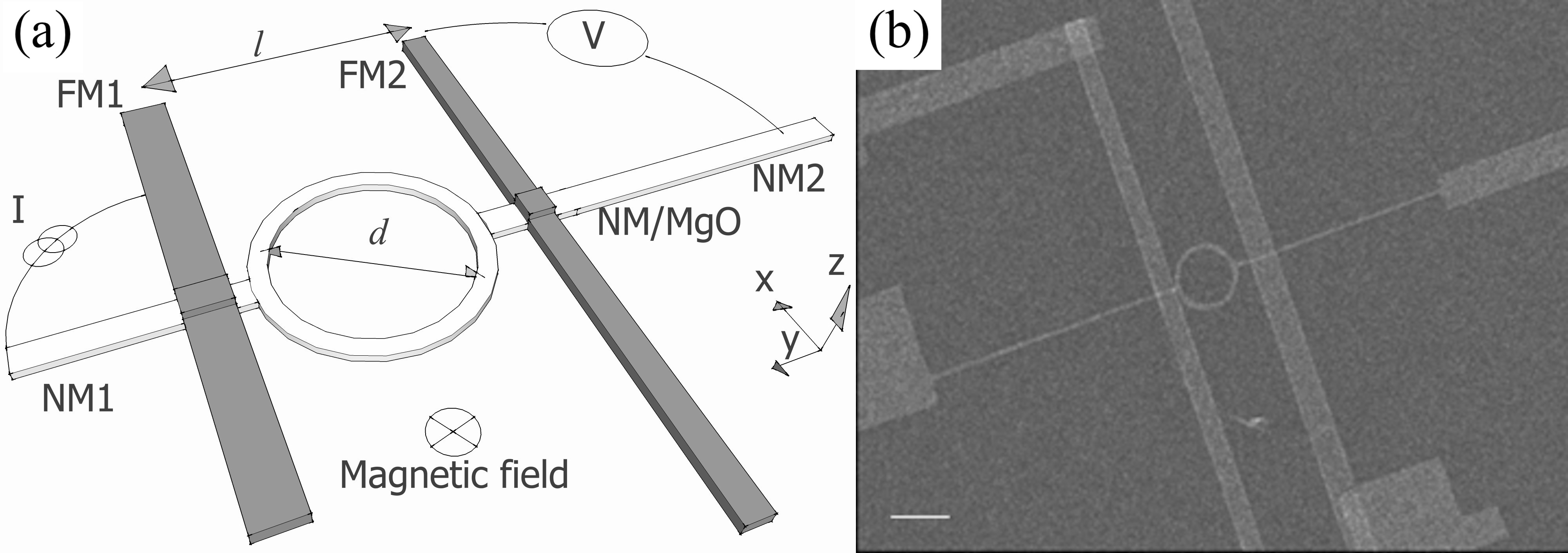}
\caption{\label{fig:sketch}(a) Sketch of the NLSV interferometer. Ferromagnetic bars FM1 and FM2 with moments parallel and in plane. AC current was applied between FM1 and NM1. With perpendicular magnetic field, the precessing spin current diffuses from FM1 to FM2 while experiencing a non-zero magnetic potential due to the magnetic flux enclosed by the ring. Non-local voltage is measured between FM2 and NM2. The distance of electron transport from FM1 to FM2 is $L = l + (\pi/2 - 1)d$. (b) SEM image of Al/MgO$\lvert$Co NLSV after electrical measurement. The scale bar is 1 $\mu$m. The diameter of the Al/MgO ring is 0.95 $\mu$m while the line width of the ring is 100 nm $\pm$ 20 nm. Different profiles of Co1 (0.4 $\mu$m wide and 18 $\mu$m long) and Co2 (0.3 $\mu$m wide and 28 $\mu$m long) give different coercive fields.}
\end{figure}

Figure \ref{fig:sketch}a shows a sketch of the interferometer. Ferromagnetic bars FM1 and FM2 serve as spin injector and detector. FM1 and FM2 spacing $l$ is relatively large to fit the ring between them. In order to get a large non-local spin signal for such a large spacing, several factors need to be considered\cite{IDZUCHI2015239}. Assuming an Elliot-Yafet mechanism, the spin-orbit coupling in combination with momentum scattering leads to spin relaxation\cite{_uti__2004}, while spin-orbit coupling is generally believed to be proportional to the fourth power of the atomic number of the metal\cite{Sinova_2015}. Therefore we use aluminum, which has the advantage of a relatively large spin diffusion length of 1.2 $\mu$m\cite{IDZUCHI2015239, Jedema:2003aa} and large phase coherence length of 2.0 $\mu$m\cite{Chandrasekhar:1985aa}. It has also been reported that a MgO tunnel barrier can enhance spin accumulation, by reducing the conductance mismatch between the FM and NM, and therefore enables long distance spin precession\cite{Fukuma_2011, Han:2010aa}. Therefore, we constructed Al/MgO$\lvert$Co(Py) NLSVs with ring-shaped NM spacers.

Figure \ref{fig:sketch}b is an SEM image of a Al/MgO$\lvert$Co device with 0.95 $\mu$m diameter ring. Two separate e-beam lithographic and deposition phases were required to build the samples. First, the NM ring and electrodes were patterned by e-beam lithography and 20 nm of Al was deposited by thermal evaporation at a base pressure of $10^{-7}$ Torr. Without breaking vacuum, 2 nm of MgO was deposited via RF magnetron sputtering. Second, e-beam lithography was again used to pattern the FM bars and, before metal evaporation, an Argon plasma was used to remove any lithographic residue and other contamination. Then, 50 nm of Co was thermally evaporated at $10^{-7}$ Torr. Finally, samples were annealed at $360\ ^\circ$C for 10 minutes at $10^{-6}$ Torr\cite{Yuasa_2004}.

For the samples considered here, the diameter $d$ of the rings, measured via SEM, were 0.95 $\mu$m and 2.2 $\mu$m. Electron transport distance $L$ was 2 $\mu$m for 0.95 $\mu$m ring. Al bar samples with $L$ of 0.5 $\mu$m, 1 $\mu$m and 2.3 $\mu$m were also fabricated at the same time to serve as controls. Al/MgO$\lvert$Co samples were measured in a cryostat with a lowest achievable temperature of 1.9 K. Another group of Py(20 nm)/MgO(2 nm)$\lvert$Al(60 nm) ring samples were fabricated and measured in a dilution refrigerator with a lowest achievable temperature of 50 mK.

\begin{figure}
\centering
\includegraphics[scale=0.045]{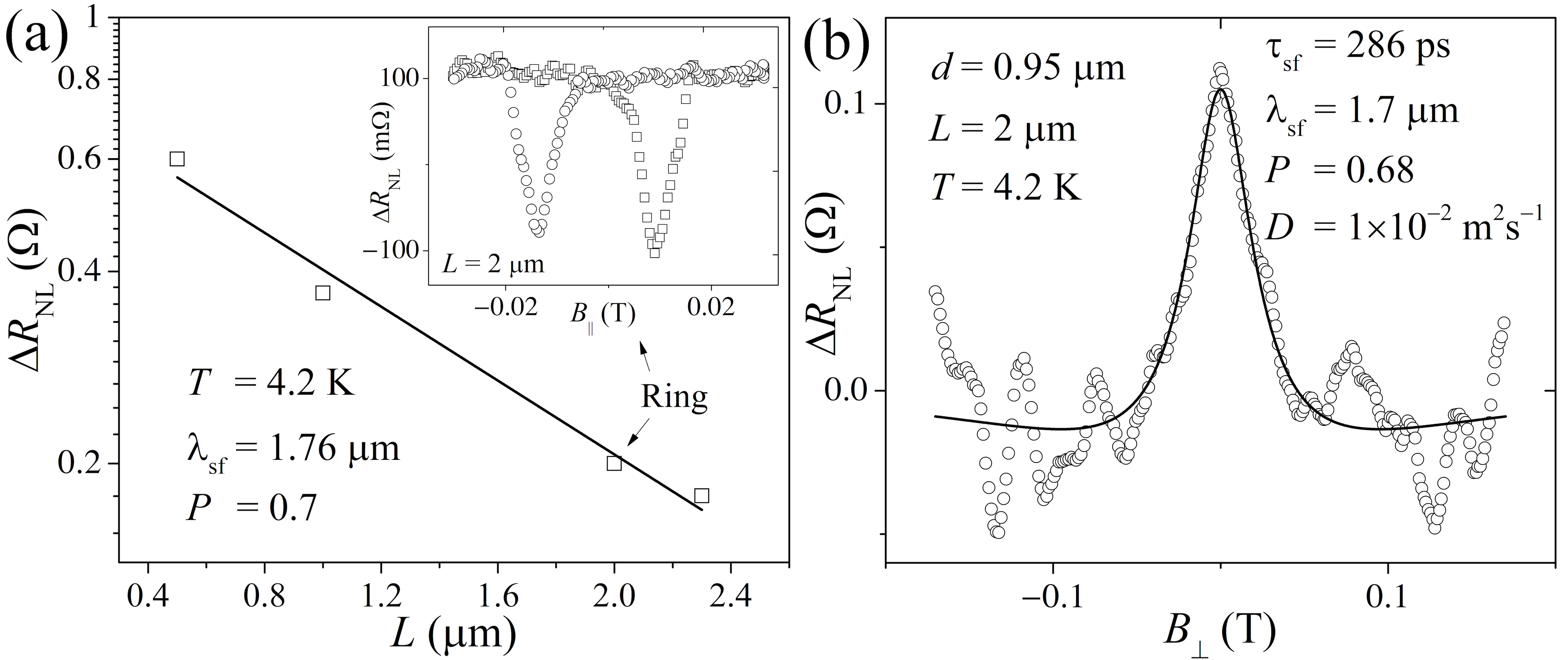}
\caption{\label{fig:dRvsLB}Spin resistance for in-plane and perpendicular fields: (a) In-plane: Distance dependence of spin resistance for both ring ($L=2$ $\mu$m) and bar samples ($L=0.5$ $\mu$m, 1.0 $\mu$m and 2.3 $\mu$m) at 4.2 K yields an average spin diffusion length of 1.76 $\mu$m and spin polarization rate of 0.7. Inset: 200 m$\Omega$ spin resistance was observed for a 0.95 $\mu$m ring with 2 $\mu$m Co1/Co2 distance. (b) Perpendicular: Black circles are data showing a clear Hanle effect curve for a 0.95 $\mu$m ring Al/MgO$\lvert$Co at 4.2 K. Solid line is a least squares fit to eq. \ref{eq:Hanle}. Only every $10^{th}$ data point is shown for clarity.}
\end{figure}

We first measured non-local voltage $V_\mathrm{NL}$ for both ring and bar samples with different $L$ using standard lock-in techniques. Alternating current, amplitude of 2 $\mu$A at 17 Hz, was applied between Co1 and NM1. $V_\mathrm{NL}$ was measured between Co2 and NM2. The in-plane field, $B_\parallel$, was swept between $\pm0.03$ T at 1 mT/s and 4.2 K. The inset of figure \ref{fig:dRvsLB}a is a typical non-local spin current measurement for the 0.95 $\mu$m ring. A very high spin resistance ($\Delta R_{\mathrm{NL}}=\Delta V_{\mathrm{NL}}/I$) of 200 m$\Omega$ was observed for $L=2\mu$m. 

Spin resistance and diffusion length are related by\cite{Jedema_2002}
\begin{equation}
\label{eq:dR}
\Delta R_{\mathrm{NL}}=\pm{\frac{1}{2}}P^2{\frac{\lambda_{\mathrm{sf}}}{{\sigma_{\mathrm{Al}}A}}}\exp(-L/\lambda_{\mathrm{sf}}).
\end{equation}
$P$ is the spin polarization of current injected into Al from Co1, $\lambda_{\mathrm{sf}}$ the spin diffusion length in the Al, and $A$ the cross sectional area of the tunnel junction. $\sigma_{\mathrm{Al}}$ is the conductivity of the Al film which was found to be $3\times10^7$ $\Omega^{-1}\mathrm{m}^{-1}$ from the linear fit in figure \ref{fig:dRvsLB}a. We find an average spin diffusion length of 1.76 $\mu$m and a spin polarization of 0.7 in both the ring and bar samples.

To determine the transverse spin diffusion length, a Hanle spin precession measurement was made with an out-of-plane magnetic field, $B_\perp$ between $\pm 0.17$ T at 1 mT/s. Figure \ref{fig:dRvsLB}b shows the resulting Hanle effect curve. The solid line fits the spin resistance to\cite{Jedema_2002}
\begin{equation}
\label{eq:Hanle}
\Delta R_{\mathrm{NL}}=\pm{\frac{P^2}{{e^2N_{\mathrm{Al}}A}}}\int_0^\infty{\mathrm{p}(t)\cos(\omega_{\mathrm{L}}t)\exp(-t/\tau_{\mathrm{sf}})\mathrm{d}t},
\end{equation}
where $\mathrm{p}(t)=(1/\sqrt{4\pi Dt})\exp{[-L^2/(4Dt)]}$ is the distribution of diffusion times from injector to detector, $e$ the electron charge, $D$ the diffusion coefficient in Al, $N_{\textrm{Al}}=2.4 \times 10^{-3}$ eV$^{-1}$cm$^{-3}$ the density of states of Al near the Fermi level\cite{Handbook}, $\omega_\mathrm{L}=g\mu_\mathrm{B}B_\perp/\hbar$ the Larmor frequency, and $g$ the g-factor in Al. From the fit, we get a spin relaxation time of $\tau_{\mathrm{sf}}=286$ ps, diffusion coefficient of 0.0108 $\mathrm{m}^2 \mathrm{s}^{-1}$ and spin polarization of 0.68. The spin diffusion length is then $\lambda_{\mathrm{sf}}=\sqrt{D\tau_{\mathrm{sf}}}=1.69$ $\mu$m, which agrees quite well with the previous distance-dependent spin resistance measurements.

The Hamiltonian for an electron with spin is
\begin{equation}
\label{eq:H}
H = {\frac{1}{2m^*}}\big[\textbf{p}+{\frac{e}{c}}\textbf{A}(\textbf{r})\big]^2+V(\textbf{R}_n)+\mu\textbf{B}\cdot\boldsymbol{\sigma}.
\end{equation}
where $m^*$ is the effective electron mass, $\textbf{p}+{\frac{e}{c}}\textbf{A}(\textbf{r})$ its generalized momentum, $\textbf{A}$ the magnetic vector potential, $\mu$ the electron magnetic moment, and $\boldsymbol{\sigma}$ the Pauli matrix. $V(\textbf{R}_n)$ is defined by the confinement of the conductor and the impurity potential\cite{Stern:1992aa, Gao:1993aa, Frustaglia:2001aa, Yi:1997aa, Hentschel:2004aa}, where $\textbf{R}_n$ is the position vector of the $n^{th}$ scattering event.

In the explanation offered in the experimental observation of AB phase in Au rings\cite{Webb}, electrons are assumed to move ballistically between elastic scattering centers $\textbf{R}_i$ and $\textbf{R}_{i+1}$. When the spin degree of freedom is not considered, $H(\textbf{R}_i)\ket{\psi_i}=E(\textbf{R}_i)\ket{\psi_i}$ holds after the scattering event at $\textbf{R}_i$ such that $\ket{\psi_i}$ is the instantaneous eigenstate of eq. \ref{eq:H} without the last term. This is how the AB phase is viewed as a special case of Berry's adiabatic geometric phase\cite{Berry_1984}. 

The complete description of the electron quantum state is $\ket{\Psi(t)}=\ket{\psi(\textbf{r},t), \textbf{S}(t)}=\ket{\psi(\textbf{r},t)}\otimes\ket{\textbf{S}(t)}$, where $\ket{\psi(\textbf{r},t)}$ is the spatial part while $\ket{\textbf{S}(t)}$ is the spin part. The injector aligns the spins in the $x$-direction (figure \ref{fig:sketch}a), i.e. $\ket{\textbf{S}(t=0)}=\ket{S_x; \pm}$, the eigenstate of $S_x$. When diffusing from injector to detector through the ring with a uniform magnetic field $\textbf{B}=B_z \hat{\textbf{z}}=B_\perp\hat{\textbf{z}}$ applied, the spin precesses in the $x$-$y$ plane at the Larmor frequency. 

\begin{figure}
\centering
\includegraphics[scale=0.04]{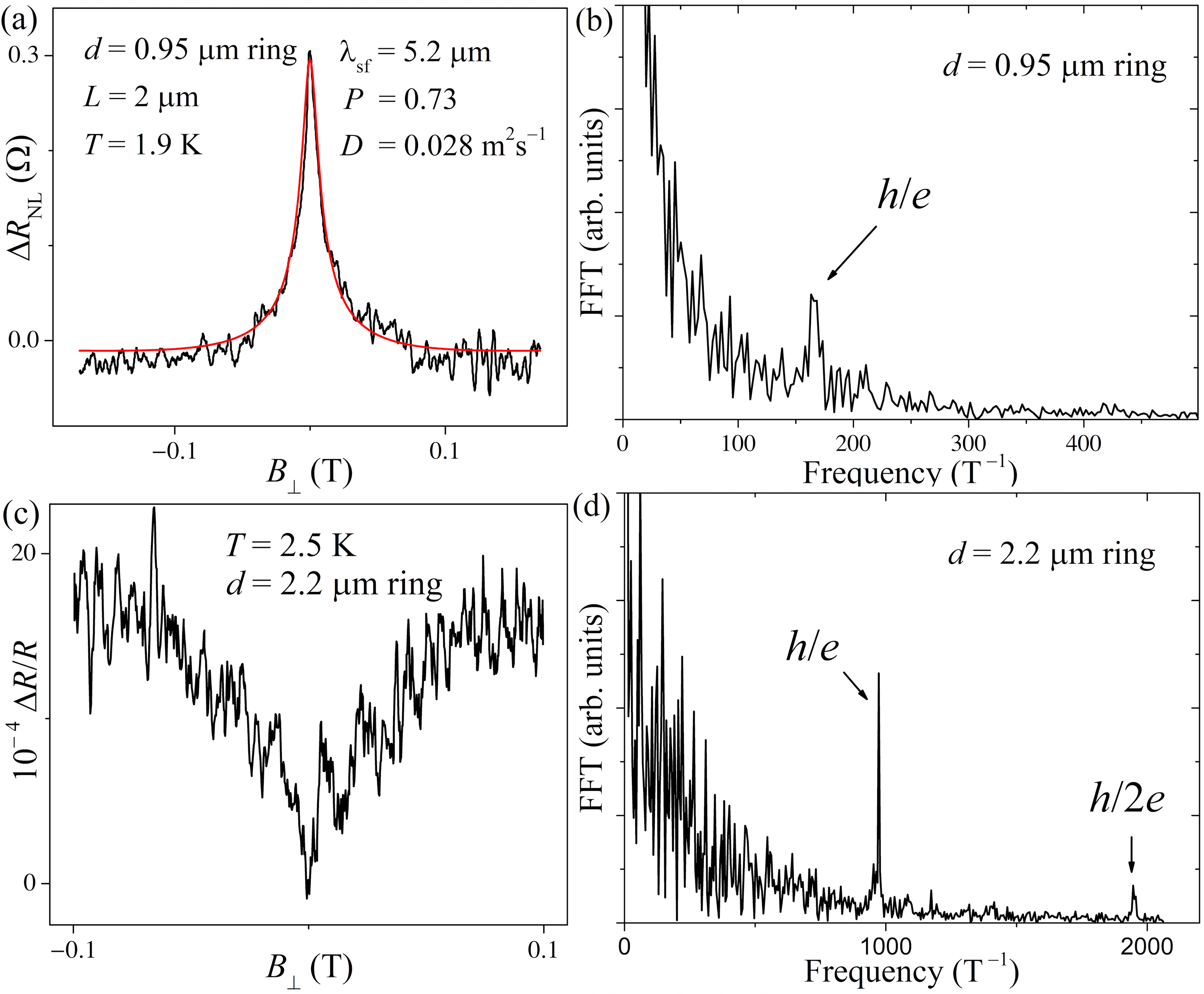}
\caption{\label{fig:fft}(a) Non-local Hanle curve for a 0.95 $\mu$m ring at 1.9 K. (b) FFT of the non-local Hanle curve shows a peak at 168 $\mathrm{T}^{-1}$, which corresponds to an $h/e$ oscillation. (c) Local Hanle effect (magnetoresistance) for a 2.2 $\mu$m ring with spin polarized current of 2 $\mu $A at 2.5 K. (d) FFT of the local Hanle curve for a 2.2 $\mu$m ring shows both $h/e$ and $h/2e$ oscillations with peaks at 973 $\mathrm{T}^{-1}$ and 1948 $\mathrm{T}^{-1}$.}
\end{figure}

The spin is perpendicular to $B_z$ after each scattering event, so the electron state ket $\Psi(t)$ is not at the instantaneous eigenstate of the Hamiltonian, eq. \ref{eq:H}, since $[H, S_z]=0$. The average value of $S_x$ can be inferred by the non-local voltage between Co2 and NM2 and the associated Hanle curve. At $B_z=0$, the electrons diffuse from injector to detector without spin precession. When $B_z$ increases to $B_{z,min}=0.15$ T as shown in figure \ref{fig:fft}a, the Hanle curve reaches its first minimum, for which the spin precession angle is $\pi$\cite{IDZUCHI2015239}. 

Since half of the ring circumference is $\frac{1}{2}\pi d$, the magnetic field which causes
 a precession of $\pi$ while diffusing a distance $\frac{1}{2}\pi d$ is $B_{z,\pi} = (2L/\pi d) B_{z,min} \approx0.19$ T. Therefore, for magnetic fields $B_z<0.19$ T, the time for an electron to transit the ring is shorter than the period of Larmor precession. Based on the above analysis, in 0.95 $\mu$m NLSV interferometer at low fields, the quantum adiabatic approximation is violated.
 
We use AA's approximate treatment to get the geometric phase factor\cite{Aharonov:1987aa},

\begin{equation}
\label{eq:Phase}
\beta=-\frac{e}{\hbar}\oint_\gamma A_\mu dx^\mu+\frac{1}{\hbar}\oint_\gamma \textbf{p}\cdot d\textbf{x}.
\end{equation}
where $\beta$ is the AA's geometric phase, $A_\mu$ the electromagnetic four-potential, $\textbf{p}$ the kinetic momentum, and $\gamma$ the space-time closed curve of two electrons propagating through the two arms for the ring. Because of time reversal symmetry between the two paths, the time component is zero, $\oint_\gamma A_0dx^0=0$. Further, $\oint_\gamma \textbf{p}\cdot d\textbf{x}=0$ for a complete trip around the ring. Therefore, we have
\begin{equation}
\label{eq:ph}
\beta=\frac{e}{\hbar}\oint_\gamma \textbf{A}\cdot d\textbf{x}=\frac{e}{\hbar}\Phi_\mathrm{B}.
 \end{equation}
where $\Phi_{\mathrm{B}}=B_z\cdot S$ is the magnetic flux through the enclosed area of the ring.

With $B_\perp$ sweeping, A sinusoidal oscillation of non-local resistance with $\Delta B_z\cdot S=\frac{2\pi\hbar}{e}$ should be observed superimposed on the standard Hanle curve because precession does not break phase coherence. Figure \ref{fig:fft}a is a Hanle effect curve for the 0.95 $\mu$m ring with AC current of 0.5 $\mu$A at 1.9 K. By taking the Fourier transform of the Hanle curve, we observe a peak at 168 $\mathrm{T}^{-1}$ as in figure \ref{fig:fft}b. This peak corresponds to a resistance oscillation with period $\Delta B_z=$ 5.9 mT. By calculating the average area of the hole enclosed by the ring, we get $\Delta B_z=$ 5.8 mT. Considering that the accuracy of area measurement is 10$\%$, this is very good agreement.

The local Hanle effect was observed by magnetoresistance measurements of a 2.2 $\mu$m diameter ring, showing an interference effect with spin polarized current and spin precession; both diffusive spin precession and drift current exist. A 2 $\mu$A AC current at 17 Hz was applied between Co1 and Co2 prepared parallel in plane and the voltage between NM1 and NM2 was measured via lock-in with $B_\perp$ sweeping. The magnetoresistance curve (Figure \ref{fig:fft}c) is weak-antilocalization-like\cite{Bergmann_1984}, but fitting to the model\cite{Hikami:1980} produces unreasonable results. This is understandable because Hikami's model does not consider spin polarization\cite{Hikami:1980}. The Fourier transform of the magnetoresistance curve is shown in figure \ref{fig:fft}d. Two peaks at 973 $\mathrm{T}^{-1}$ and 1946 $\mathrm{T}^{-1}$ were clearly observed and, from equation \ref{eq:ph} with $\beta = 2\pi$, correspond to $h/e$ and $h/2e$ oscillations, respectively.

Previous works\cite{Poli:2008aa,Miura_2006} on NLSV bar systems have shown a marked change in $\Delta R_\mathrm{NL}$ as the temperature crosses $T_c$ from above, therefore, we made non-local spin measurements at a range of temperatures from 50 milliKelvin to 1.6 K, above $T_c$. A Py/MgO$\lvert$Al 1 $\mu$m diameter ring sample was cooled to 50 mK in a dilution refregerator. Non-local resistance was measured with an AC injection currents of 0.5 $\mu$A (quasi-static measurements) and 2 $\mu$A (swept field measurements) at 79 Hz\cite{Fukuma_2011}. The inset of figure \ref{fig:noquasi}a shows $T_c$ detection with 3 different FM configuration. We find $T_c=1.54$ K with the non-local resistance baseline jumping from 127.6 $\Omega$ to 122 $\Omega$. $T_c$ shift of 5 mK, which is due to spin polarization difference\cite{Miao:2007aa}, was observed only with opposite magnetization direction of injector. Spin current is clearly seen below, above, and at $T_c$ in figure \ref{fig:noquasi}b. The critical field at $T=1.5$ K was found to be $B_{c, \parallel}=0.025$ T.

To compare the spin resistance, we offset the 1.6 K data as in figure \ref{fig:noquasi}b. $\Delta R_\mathrm{NL}$ in the superconducting state is almost identical to that in the normal state. However, below $T_c$, as the field approaches the point when either FM1 and FM2 are in the negative parallel configuration, the non-local resistance increases (arrows in figure \ref{fig:noquasi}b). The magnitude of the increase grows as the temperature approaches $T_c$ from below and vanishes above $T_c$. This effect may come from the interaction between Cooper pairs and magnetic domains\cite{Linder_2015, Linder:2014aa}. 

The almost identical spin resistance below and above $T_c$ is inconsistent with theoretical predictions based on the quasiparticle picture\cite{Takahashi:2003aa} which suggest a temperature dependent enhancement of spin resistance by the factor of $1/2f_0(\Delta)$ at the superconducting state, where $\Delta$ is the superconducting gap and $f_0$ is the Fermi distribution function\cite{Takahashi:2003aa, Poli:2008aa, Beckmann_2016}. The magnitude of the enhancement was predicted to be several orders of magnitude above the normal state\cite{Takahashi:2003aa, Poli:2008aa, Beckmann_2016}.

The co-existence of superconductivity and spin current in our device could be direct evidence of spin supercurrents from spin triplet Cooper pairs, which were under experimental and theoretical investigation recently\cite{Linder_2015}. Because we continue to observe spin current in the superconducting state, we suggest that, similar to superfluid $^3$He, the spin triplet states $\ket{\uparrow\uparrow}$, $\ket{\downarrow\downarrow}$ and $\frac{1}{\sqrt{2}}(\ket{\uparrow\downarrow}+\ket{\downarrow\uparrow})$, among which $\ket{\uparrow\uparrow}$ and $\ket{\downarrow\downarrow}$ have non-zero spin and therefore can also cause a spin current, may also exist in thin films of Al with direct spin injection from a FM, even though Al is traditionally categorized as a type $\mathrm{\Romannum{1}}$ superconductor in which only spin singlet Cooper pairs exist\cite{Wheatley:1975aa, Balatsky:2006aa, Sigrist:1991aa}. Consistent with this supposition, recent experimental work on spin pumping into superconductors has shown evidence of triplet spin supercurrent and not quasiparticles\cite{Jeon:2018}. For a clear understanding of the spin transport behavior below $T_c$, more theoretical and experimental work is needed.

To conclude, we experimentally observed simultaneous h/e oscillation and Hanle spin precession in the NLSV interferometer above $T_c$ for an Al ring spacer. Because the non-adiabatic condition is satisfied at low fields when the time for an electron to transit the ring is shorter than the period of Larmor precession, our measurement is an experimental observation of AA's non-adiabatic geometric phase and not an AB phase measurement. In addition, we observed the co-existence of spin current and superconductivity. The identical spin resistance above and below $T_c$ is inconsistent with predictions from quasiparticle theories; we suggest the superconducting spin current may come from spin triplet Cooper pairs.

\begin{acknowledgments}
We thank Professor Richard A. Webb for initial discussions, Professor Thomas M. Crawford for help while making measurements at the National High Magnetic Field Laboratory, and Professor Scott R. Crittenden for helpful discussions. This work was supported by the University of South Carolina SmartState Center for Experimental Nanoscale Physics. A portion of this work was performed at the National High Magnetic Field Laboratory High B/T facility, which is supported by National Science Foundation Cooperative Agreement No. DMR- 1644779, the State of Florida, and the U.S. Department of Energy.
\end{acknowledgments}

\end{document}